\documentclass[useAMS,usenatbib]{mn2e}
\usepackage{epsfig}

\newcommand{\kms}{\mbox{km\,s$^{-1}$}}

\def\ms{\,m\,s$^{-1}$}         %m.s -1
\def\kms{\,km\,s$^{-1}$}       %km.s -1
\def\kms{km\, s$^{-1}$}

\title[Spin-orbit misalignement for the transiting planet HD~80606b]{Spin-orbit misalignement for the transiting planet HD~80606b}

\author[M. Gillon]{M. Gillon$^{1,2}$\thanks{E-mail:
michael.gillon@ulg.ac.be} \\  
$^1$ Institut d'Astrophysique et de G\'eophysique,  Universit\'e
  de Li\`ege,  All\'ee du 6 Ao\^ut, 17,  Bat.  B5C, Li\`ege 1, Belgium \\
$^2$  Observatoire de Gen\`eve, Universit\'e de Gen\`eve, 51 Chemin des Maillettes, 1290 Sauverny, Switzerland}
\begin{document}

\date{Received date / accepted date}

\pagerange{\pageref{firstpage}--\pageref{lastpage}} \pubyear{2009}

\maketitle

%%%%%%%%%%%%%%%%%%%%%%%%%%%%%%%%%%%%%%%%%%%%%%%%%%%%
\begin{abstract}
A global Markov Chain Monte Carlo analysis of published eclipse photometry and radial velocities is presented for the transiting planet HD~80606b. Despite the lack of a complete transit light curve, the size of the planet is measured with a good level of precision ($R_p = 1.04^{+0.05}_{-0.09}$ $R_{Jup}$),  while the orbital parameters  are refined. This global analysis reveals that the orbital axis of the planet is significatively  inclined relative to the spin axis of the host star ($\beta = -59^{+18}_{-28}$ deg), providing a compelling evidence that HD~80606b owes its peculiar orbit to the Kozai migration mechanism.  
\end{abstract}

\begin{keywords}
binaries: eclipsing -- planetary systems -- stars: individual: HD~80606 -- stars: individual: HD~80607  -- techniques: photometric  -- techniques: spectroscopic -- methods: data analysis
\end{keywords}
%%%%%%%%%%%%%%%%%%%%%%%%%%%%%%%%%%%%%%%%%%%%%%%%%%%%

\section{Introduction}

Among the known transiting gazeous planets, HD~80606b is definitely an unique case. Detected by radial velocity measurements (Naef et al. 2001), this massive planet ($M \sim 4$ $M_J$) orbits in a very eccentric orbit ($e \sim 0.93$) around one of the two solar-type components of a wide binary ($\sim$ 1000 AU). With $P \sim 111$ days, it is by far the known transiting planet with the largest period. Laughlin et al. (2009; herafter L09) observed this planet with $Spitzer$ during its periastron passage and detected its occultation, implying an inclination close to 90$^{\circ}$.  A partial photometric transit was firmly detected a few months later by several ground-based telescopes (Fossey et al. 2009, hereafter F09; Garcia-Melendo \& McCullough 2009, hereafter G09; Moutou et al. 2009, hereafter M09). This transit was also detected by M09 in spectroscopy through the observation of the Rossiter-McLaughlin effect (Queloz et al. 2000).

All three transit detection papers report a size close to Jupiter's for the planet. Still,  the shape and duration of the transit are poorly constrained by the ground-based partial transit data. G09 considered a grazing transit  as unlikely, basing on the similar transit depths  measured in $B$-band (G09) and $R$-band (F09, M09), but did not consider the possibility of an imperfect normalization or the presence of a trend or low-frequency noise (due, e.g., to an imperfect colored extinction correction or to a low-frequency variability of the star) in one or more of these ground-based time-series. Assessing the influence of such an effect is especially important here as it could improve largely the probability of a grazing transit, and thus of a larger planet. To investigate this point and to assess the validity of the reported planetary size, a global analysis of all the data available for the HD~80606 system was performed, considering possible the presence of  trends in the eclipse photometry. This analysis presented here aimed also  to obtain a reliable estimate of the significance of the possible spin-orbit misalignement suggested in M09. The confirmation of such a misalignement for HD~80606b would be a strong argument for the hypothesis that the planet owes its peculiar present orbit to the Kozai migration mechanism (Wu \& Murray 2003). 

The data used in this work and a new photometric reduction of the $Spitzer$ images (L09) are presented in Sec. 2. The analysis itself is described in Sec. 3. In Sec. 4, the results of the analysis and some of their implications are discussed.

\section{Data}

The global determination of the system parameters was performed using the published radial velocities obtained by the spectrographs ELODIE (Naef et al. 2001; M09), SOPHIE (M09), HRS (Wittenmeyer et al. 2007) and HIRES (Butler et al. 2006, L09), the transit photometry obtained in $B$-band by G09 and the one obtained in $R$-band by M09 and F09. It has to be noticed that the publically available M09 photometry\footnote{http://cdsarc.u-strasbg.fr/viz-bin/Cat?J/A\%2bA/498/L5} presents unrealistic error bars, and  an extra noise of 0.25 \% has to assumed to reproduce the $rms$ of the best-fitting solution. 

The occultation observed by L09 brings an important constraint on the orbital solution. Instead of using its published timing (L09) as a prior knowledge in the analysis, it was chosen to model the L09 $Spitzer$ photometry together with the other data. Indeed, an indirect but important constraint on the transit shape comes from the $occultation$ shape, the parameters of the  two eclipses being of course correlated. This is important here, because of the lack of a full transit. The $Spitzer$ IRAC occultation  photometry presented in L09 shows a significant level of correlated noise, but this latter is mostly due to the `per-pixel' photometric reduction done in that study to isolate reliably the heating of the planet during its periastron passage (D. Deming, priv. comm.). A more classical reduction of these data was thus performed to better constrain the shape of the occultation. Starting from the   IRAC images calibrated by the standard $Spitzer$ pipeline and delivered to the community as Basic Calibrated Data (BCD),  the fluxes were converted from the $Spitzer$ units of specific intensity (MJy/sr) to photon counts, and aperture photometry was obtained for HD~80606 and HD~80607 in each image using the {\tt IRAF/DAOPHOT}\footnote{ {\tt IRAF} is distributed by the National Optical Astronomy Observatory, which is operated by the Association of Universities for Research in Astronomy, Inc., under cooperative agreement with the National Science Foundation.} software (Stetson, 1987). The aperture was centered in each image by fitting a Gaussian profile on the targets. A mean sky background was measured in an annulus extending from 10 to 20 pixels from the center of the aperture, and subtracted to the measured fluxes for each image. Each flux measurement was compared to the median of the 10 adjacent images and rejected as outlier if the difference was larger than 4 times its theoretical error bar ($\sim$ 0.16 \%). The aperture was adjusted so as to obtain the highest quality on the lightcurve of HD~80607, an aperture of 3.5 pixels giving the best result. Finally, only 1709 measurements  ($\sim$ 6.5h)  encompassing the occultation were kept to make possible the modelisation of the combined effect of the heating modulation and the IRAC systematic known as the `ramp' (see e.g. Knutson et al. 2008 and references therein) by a simple analytical function. Fig.~1 (first panel from the top) shows the resulting occultation photometry. 

\section{Global analysis}

The global analysis was performed with a modified version of the Markov chain Monte Carlo (MCMC) code described in Gillon et al. (2009). MCMC is a Bayesian inference method based on stochastic simulations and provides the posterior probability distribution of adjusted parameters for a given model (see e.g. Tegmark et al. 2004, Gregory 2005, Ford 2006).  The Metropolis-Hasting algorithm was here introduced into a Gibbs sampler to improve the convergence speed of the chains (see e.g. Ford 2006). For all the data, a normal distribution was assumed for the errors. The used model was based on a star and a transiting planet on a Keplerian orbit about their center of mass. More specifically,  a classical Keplerian model was used for the radial velocities obtained outside the transit in addition to a Rossiter-McLaughlin effect model (Gim\'enez 2006) for the SOPHIE measurements obtained during transit, while the eclipse model of Mandel \& Agol (2002) multiplied by a trend/systematic model was used for the different eclipse photometric time-series.  For both the $B$-band and $R$-band time-series, a time-dependent quadratic polynomium was used to model the trend. A similar polynomium was also enough to model the flux variation outside the eclipse in the IRAC photometry, more sophisticated models did not improve the fit. 

The jump parameters of the MCMC analysis were: the planet to star radius ratio $R_p/R_\ast$, the parameter $b' = a \cos i /R_\ast$ ($a$ being the semi-major axis and $i$ the orbital inclination), the transit duration $W$ (first to fourth contact), the transit time of minimum light $T_0$, the orbital period $P$, the Lagrangian parameters $e\cos\omega$ and $e\sin\omega$ ($e$ being the eccentricity and $\omega$ the argument of periastron), $K_2 = K \sqrt{1-e^2} P^{1/3}$ ($K$ being the RV orbital semi-amplitude), the products $V\sin I \cos\beta$ and $V\sin I \sin\beta$ ($V\sin I$ being the projected rotational velocity of the star and $\beta$ the spin-orbit angle, see Gim\'enez 2006), 3 trend coefficients for both transit time-series and for the IRAC photometry, a different systemic radial velocity for each spectrograph, and finally the  8 $\mu$m occultation depth $dF_2$. For the transit time-series, a quadratic limb darkening law was asssumed, with coefficients interpolated from Claret's tables (2000; 2004) for both filters and for the spectroscopic parameters derived in M09. These coefficients were kept fixed in the analysis.

 Due to is Bayesian nature, MCMC allows the  easy use of priors based on external measurements/constraints. It is a major advantage for this analysis, as no complete high-quality transit light curve is available to determine unambigously the transit parameters. The different priors used in the analysis were: \begin{itemize}
\item $V \sin I$ was determined spectroscopically by different teams (Naef et al. 2001, Valenti \& Fisher 2005). These results were combined and the result $V\sin{I} = 1.4 \pm 1.0$ km s$^{-1}$ was used  in a Bayesian penalty added to the merit function of the MCMC.  
\item Basing on a compilation of spectroscopic data from the literature and on their stellar evolution modeling, M09 derived a value $M_\ast = 0.98 \pm 0.10$ $M_{\odot}$ for the stellar mass and $R_\ast = 0.98 \pm 0.07$ $R_{\odot}$ for the stellar radius. The resulting stellar density $\rho_\ast = 1.04 \pm 0.25$ $\rho_\odot$ was used here in a Bayesian penalty added to the merit function. The stellar density itself is not a jump parameter in the MCMC, but is derived at each step from several jump parameters. On its side, the stellar mass was allowed to vary normally within its error bar to propagate its uncertainty to the other physical parameters. 
\item HD~80606 was photometrically monitored from Arizona by the MEarth project (Irwin et al. 2009) the night before the detection of the transit from Europe, until $\sim$ 2454876.00 HJD. J. Irwin reported an absence of  ingress in the resulting light curve (G. Laughlin's web site oklo.org\footnote{http://oklo.org}). Using the time sampling and scatter presented in the corresponding oklo.org post and assuming an absence of correlated noise, these MEarth  data were simulated and the resulting time-series was added as input data to the analysis, the goal being to give a correct statistical weight to this piece of information instead of rejecting $a$ $priori$ any solution with a transit beginning before 2454876.00 HJD. 
\end{itemize}

MCMC is not a global optimization algorithm, in the sense that it aims to quantify the uncertainties of a given solution and not to  fully explore the whole model parameters hyperspace to locate several well-separated merit function minima. To perform in a first stage such a global exploration, a grid of 1000 different starting points sampling the part of the parameters hyperspace close to the published orbital solutions (F09, G09, L09, M09) was set up. For each cell of the grid, a  short MCMC (1000 iterations) was performed. From the merit function minima obtained for each grid, the existence of a large and continuous minimum area was noticed.  Outside this area,  many chains managed to fit well the radial velocities and the partial transits, succeeded to predict the occultation at the observed timing, but failed to reproduce the $duration$ of the occultation. A few chains succeeded a good fit of the whole set of data but leaded to an unrealistic stellar density (and thus to a larger merit function), illustrating the interest to use as many priors as possible to avoid solution degeneracies. 

At this stage, a MCMC of 10$^{6}$ steps starting from a random location within the minimum area was launched. The residuals of the deduced best-fitting solution were then analyzed to determine the jitter noise of the radial velocities and the red noise of the photometric time-series (see Gillon et al. 2009). The deduced values  were used to update the error bars of the corresponding measurements. 10 new chains (of $5 \times 10^{5}$ iterations each) starting from different locations sampling the solution area were then performed. The first 20\% of each chain were discarded, then the 10 chains were  combined, their convergence being checked succesfully with the  Gelman and Rubin statistic test (Gelman \& Rubin 1992). Table 1 gives the resulting best-fitting value, and the 1-$\sigma$ (68.3 \%) and 3-$\sigma$ (99.7 \%) error bars for  each jump and physical parameter, while Fig. 1 shows the best-fitting model for the eclipses and orbital radial velocity variations.

As can be seen in Table 1, a grazing transit is rejected to a high level of confidence, and the size of the planet is thus rather well constrained. Interestingly, a  zero value for $\beta$ is  also rejected to a very high level of confidence. It can also be noticed that the combined analysis of all the available data leads to a significatively refined orbital solution. 

\section{Discussion}

The global analysis presented here leads to the secure conclusion that  the orbital axis of HD~80606b and the spin axis of its host star are misaligned. Such a misalignement was detected only for the massive planet XO-3b until now (H\'ebrard et al. 2008, Winn et al. 2009). A migration mechanism based on planet-planet scattering could explain the misalignment of XO-3b (see Winn et al. 2009 and references therein) but not the one of the extremely eccentric HD~80606b (Wu \& Murray 2003). These last authors presented another migration scenario to explain the peculiar orbit of HD~80606b. Under this so-called Kozai migration scenario, HD~80606b formed at a few AU from its host star, within a protoplanetary disk very inclined relative to the stellar binary plane. HD~80607 would have then induced large eccentricity and inclination variations via the Kozai mechanism (see Mazeh et al. 1997 and references therein). During the episodes of large eccentricity, tidal forces induced by HD~80606 would have been large enough to induce an inward migration of the planet, finally taking it out of the Kozai cycles. From then, the orbital evolution of the planet would have been totally dominated by tidal circularization.  The spin-orbit misalignement deduced here is in good agreement with this Kozai migration mechanism, as the Kozai cycles should have left the planet in a highly inclined orbit. 

Most of the known transiting planets have a very short period making possible getting a large amount of good observational constraints on their orbital parameters within a few lapse of time. For these planets, the observation of a partial transit can for instance be discarded from the analysis. In the case of a `long' period planet like HD~80606b, every piece of data  has to be exploited to make possible scientific inferences in the first stages of the system characterization. The Bayesian method MCMC is very well suited for such global analysis as shown by the refinement of the orbital parameters obtained for HD~80606b in the present analysis. Despite the partial nature of the only transit observed so far and the possible presence of low-frequency correlated noise in the ground-based photometry,  the planetary radius reveals to be already well constrained by the existing data and to be in concord with basic models of irradiated planets (e.g. Burrows et al. 2007, Fortney et al. 2007). It is worth mentionning that a preliminary MCMC analysis similar to the one presented above but  performed $without$ the 2 $R$-band light curves presented in F09 led to a much larger uncertainty on the planetary radius ($R_p = 1.1^{+0.4}_{-0.1}$ $R_{Jup}$), because a good fit of the data could also be obtained with a grazing transit solution. It is no more the case when the F09 data are added to the analysis, demonstrating the interest of a global analysis to maximize the observational constraints. 

Getting a precise full transit light curve of HD~80606b  and more spectroscopic transit data is of course desirable to refine the planetary density and spin-orbit misalignement. The measurement of the planetary thermal emission and global heating at other wavelenghts than 8 $\mu$m is also desirable to improve our understanding of the atmospheric properties of this fascinating extrasolar planet. 
 
\begin{figure}
\label{fig:d}
\centering                     
\includegraphics[width=8cm]{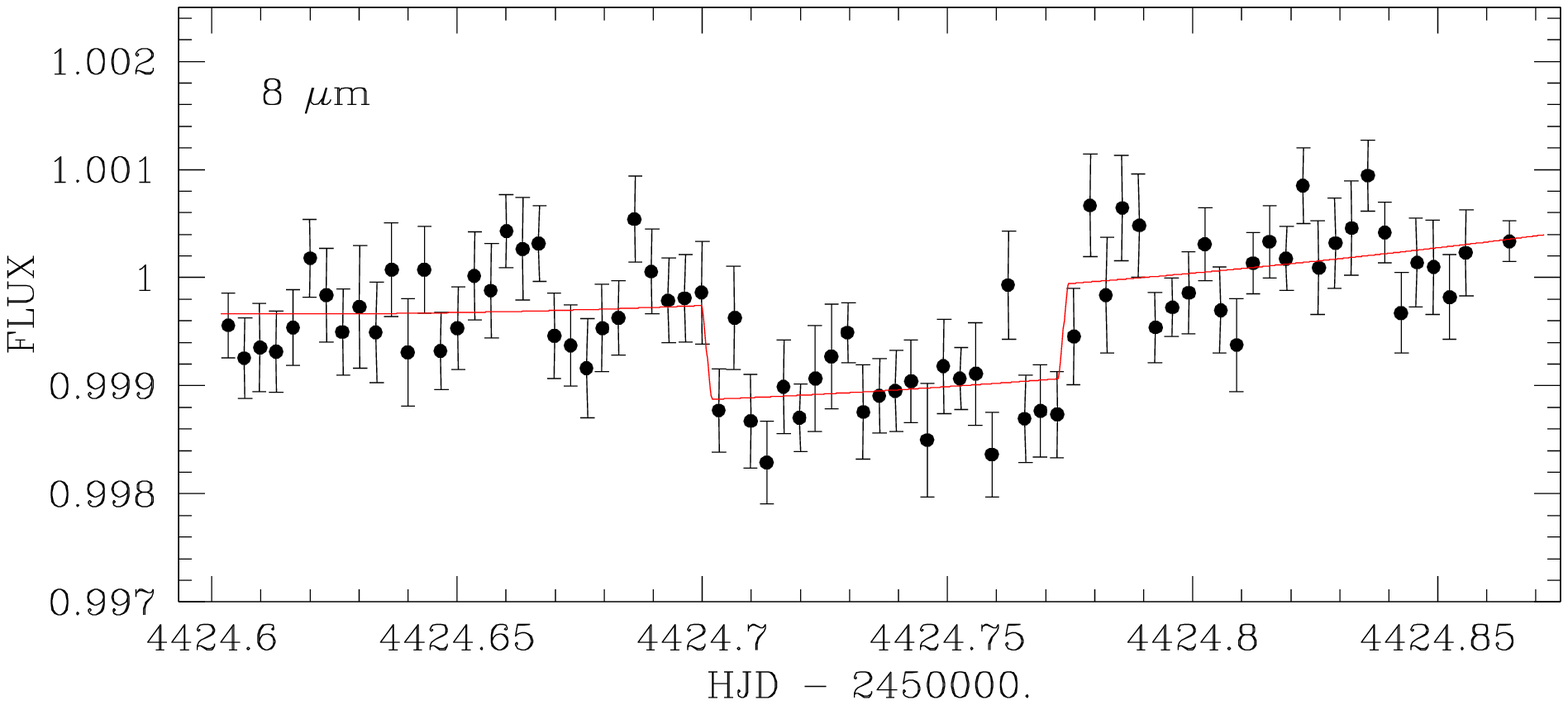}
\includegraphics[width=8cm]{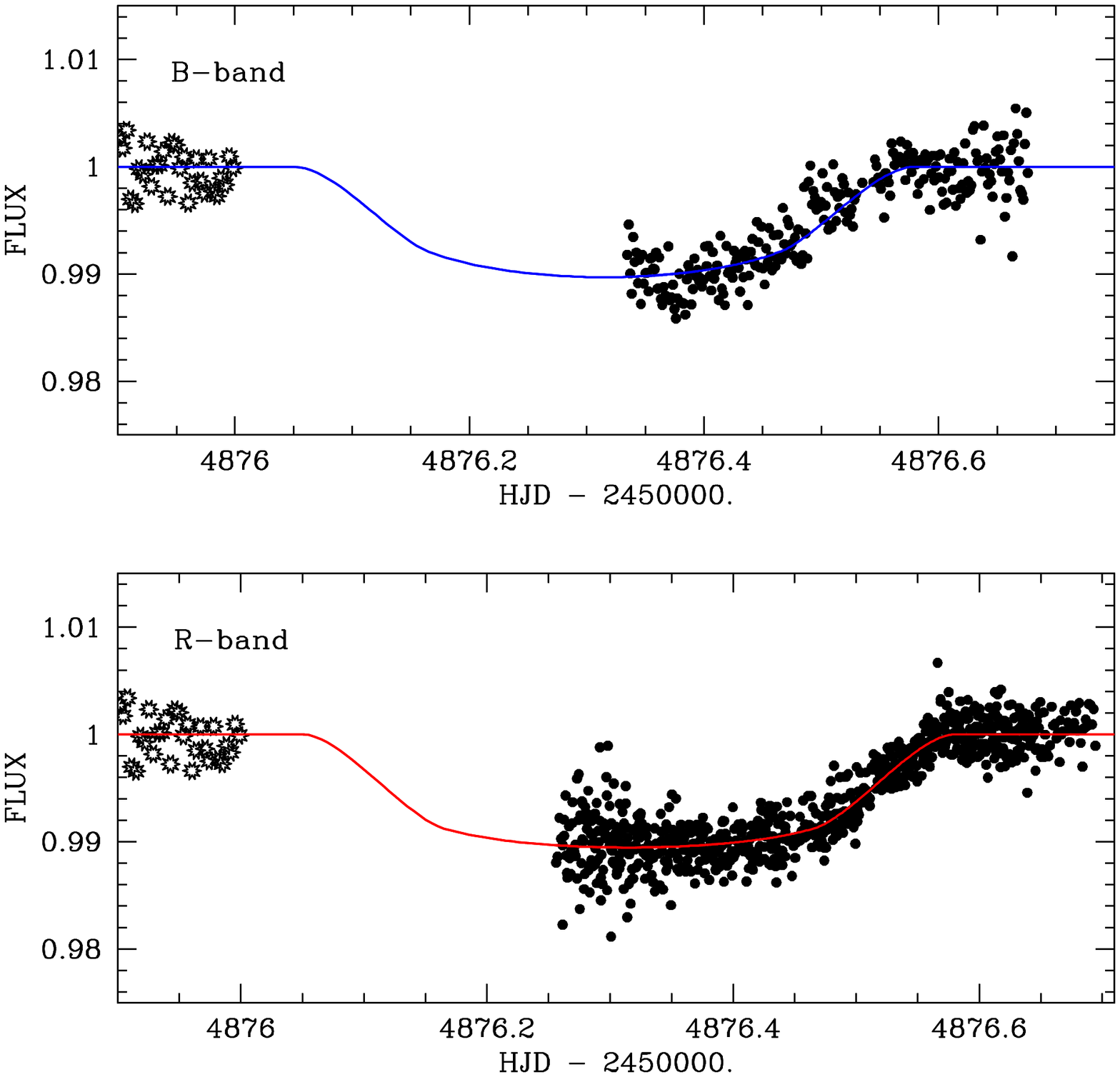}
\includegraphics[width=8cm]{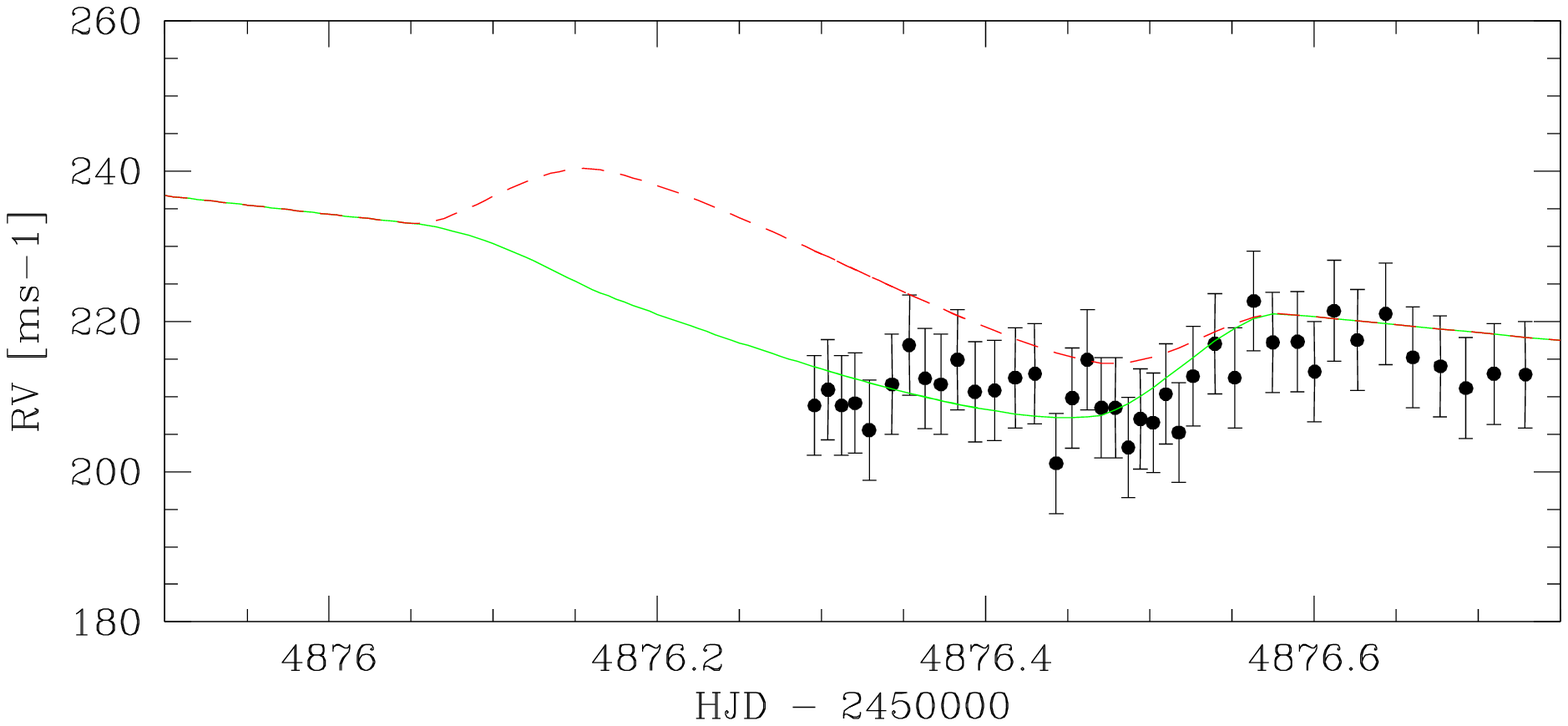}
\includegraphics[width=8cm]{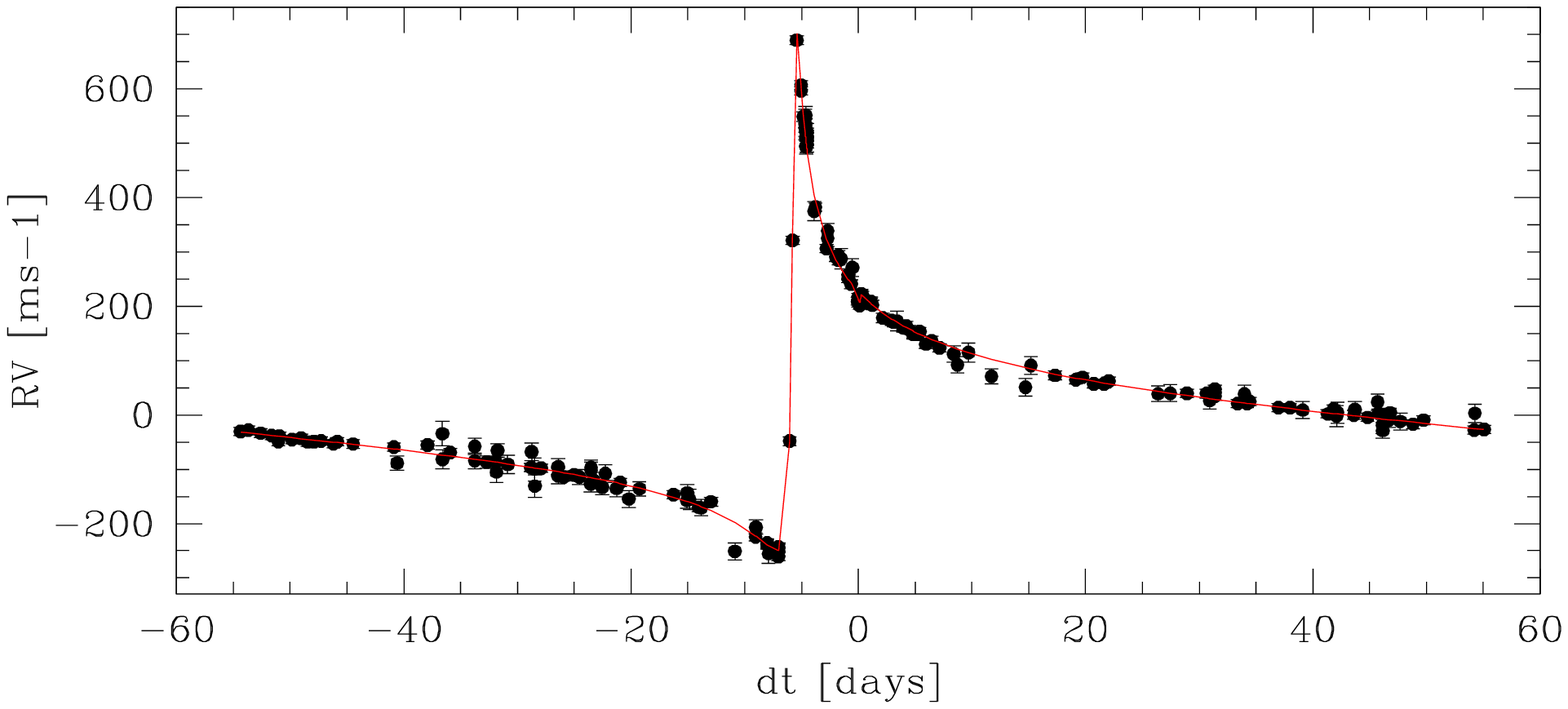}
\caption{From top to bottom: (1) IRAC 8 $\mu$m occultation photometry (binned per 5 minutes) with the best-fitting model superimposed. $B$-band (2) and  $R$-band (3) transit photometry divided by the best-fitting trend model and with the best-fitting transit model superimposed. The open symbols denote the synthetic MEarth data. (4) Radial velocities obtained by SOPHIE during the transit with 2 models superimposed: the best-fitting one (green line) and a model assuming $\beta$ = 0 deg (red dashed line). (5) The whole set of radial velocities used in this analysis folded using the best-fitting transit ephemeris.}
\end{figure}

\begin{table}
\label{tab:params}
\begin{tabular}{lcccl}
\hline
Parameter  & Value   & Units \\ \noalign {\smallskip}
\hline \noalign {\smallskip}
 $(R_p/R_s)^2 $ & 0.0106$^{+0.0008}_{-0.0003}$   \big[$^{+0.0014}_{-0.0005}$\big]  &  \\ \noalign {\smallskip}
 $ b'=a\cos{i}/R_\ast $ &  1.182$^{+ 0.086}_{- 0.001}$   \big[$^{+ 0.117}_{- 0.016}$\big]& $R_*$  \\ \noalign {\smallskip}
$W$ & 0.526$^{+ 0.003}_{- 0.072}$   \big[$^{+ 0.008}_{- 0.088}$\big]  &   days  \\ \noalign {\smallskip}
$T_0$ & 2454876.323$^{+0.035}_{-0.002}$  \big[$^{+0.042}_{-0.004}$\big]& HJD  \\ \noalign {\smallskip}
$ P$ & 111.43637$^{+ 0.00034}_{- 0.00034}$  \big[$^{+ 0.00094}_{- 0.00084}$\big]&  days  \\ \noalign {\smallskip}
$e\cos{\omega}$ & 0.4760$^{+0.0004}_{-0.0013}$   \big[$^{+0.0006}_{-0.0015}$\big]  & \\ \noalign {\smallskip}
$e\sin{\omega}$  & -0.80322$^{+0.00070}_{-0.00081}$  \big[$^{+0.00085}_{-0.00096}$\big]& \\ \noalign {\smallskip}
$V\sin{I}\cos{\beta}$ & 0.70$^{+0.27}_{-0.64}$  \big[$^{+0.65}_{-1.22}$\big] & \\ \noalign {\smallskip}
$V\sin{I}\sin{\beta}$ & -1.16$^{+0.34}_{-0.10}$  \big[$^{+0.67}_{-0.23}$\big] & \\ \noalign {\smallskip}
8 $\mu$m $dF_2$ & 0.00087$^{+ 0.00010}_{- 0.00009}$    \big[$^{+ 0.00017}_{- 0.00015}$\big]&  \\ \noalign {\smallskip}
$K_2$ & 820.8$^{+3.1}_{-3.4}$  \big[$^{+9.2}_{-8.5}$\big]&  \\ \noalign {\smallskip}
& &  \\
$V\sin{I}$ & 1.36$^{+0.08}_{-0.44}$  \big[$^{+0.27}_{-0.83}$\big] & \kms \\ \noalign {\smallskip}
$\beta$ &  -59$^{+18}_{-28}$  \big[$^{+28}_{-62}$\big] & degrees \\ \noalign {\smallskip}
$K$ &  476.2$^{+1.4}_{-2.2}$   \big[$^{+5.0}_{-5.8}$\big] &  \ms \\ \noalign {\smallskip}
$b_{transit}$ &    0.771$^{+0.059}_{-0.006}$  \big[$^{+0.080}_{-0.011}$\big] &  $R_*$ \\ \noalign {\smallskip}
$b_{occultation}$ &  0.0841$^{+ 0.0063}_{- 0.0008}$   \big[$^{+0.0089}_{-0.0012}$\big] &  $R_*$ \\ \noalign {\smallskip}
$a $ &  0.458$^{+0.012}_{-0.020}$  \big[$^{+0.035}_{-0.056}$\big] & AU \\ \noalign {\smallskip}
$a/R_\ast$ &  96.8$^{+8.5}_{-0.9}$  \big[$^{+13.5}_{-2.3}$\big] & \\ \noalign {\smallskip}
$i $ &   89.300$^{+0.029 }_{-0.046}$ \big[$^{+0.054}_{-0.078}$\big] & degrees \\ \noalign {\smallskip}
$e $ &  0.93366$^{+0.00007}_{-0.00037} $  \big[$^{+0.00014}_{-0.00043}$\big] &    \\ \noalign {\smallskip}
$\omega $ &  300.651$^{+0.041}_{-0.097}$ \big[$^{+0.058}_{-0.109}$\big] &  degrees  \\ \noalign {\smallskip}
$M_\ast $ & 1.03$^{+0.08}_{-0.13}$ \big[$^{+0.25}_{-0.33}$\big] &    $M_\odot$  \\ \noalign {\smallskip}
$R_\ast $ & 1.017$^{+ 0.026}_{- 0.093}$  \big[$^{+ 0.077}_{- 0.197}$\big] &  $R_\odot$ \\ \noalign {\smallskip}
$\rho_* $ & 0.98$^{+ 0.28}_{- 0.03}$  \big[$^{+ 0.47}_{- 0.07}$\big] &  $\rho_\odot $\\ \noalign {\smallskip}
$M_p $ & 4.12$^{+ 0.21}_{- 0.35}$  \big[$^{+ 0.63}_{- 0.94}$\big] &   $M_{Jup}$ \\ \noalign {\smallskip}
$R_p $ & 1.043$^{+ 0.048}_{- 0.087}$ \big[$^{+ 0.121}_{- 0.185}$\big] &  $R_{Jup}$ \\ \noalign {\smallskip}
$\rho_p $ & 3.63$^{+0.93}_{-0.41}$ \big[$^{+1.78}_{-0.72}$\big] &   $\rho_{Jup}$ \\ \noalign {\smallskip}
\hline\\ \noalign {\smallskip}
\end{tabular}
\caption{ Best-fitting values and 68.3\% and 99.7\% confidence intervals deduced from the marginalized posterior distribution for the jump and physical parameters.}
\end{table}

\section*{Acknowledgements} 

The author acknowledges support from the Belgian Science Policy Office in the form of a Return Grant, and thanks E. Garcia-Melendo, S. J. Fossey, D. M. Kipping, P. R. McCullough , G. Laughlin and I. P. Waldman for having provided him their data. F. Bouchy and B.-O. Demory are also thanked for interesting discussions and comments about the planet HD~80606b.

\end{document}